\documentclass[10pt]{article}
\usepackage[letterpaper,margin=0.75in]{geometry}
\usepackage{setspace}
\usepackage{amsmath,amssymb,amsthm,mathtools}
\usepackage{authblk}
\usepackage{booktabs,tabularx,array,threeparttable,longtable}
\usepackage{pdflscape}
\usepackage{placeins}
\usepackage[round,authoryear,sort]{natbib}
\usepackage{hyperref}
\hypersetup{hidelinks}
\usepackage{bm}
\usepackage{graphicx}

\newtheorem{theorem}{Theorem}[section]

\newtheorem{corollary}[theorem]{Corollary}

\numberwithin{equation}{section}

\newcommand{\E}{\operatorname{E}}
\newcommand{\Pbb}{\mathbb{P}}
\newcommand{\Var}{\operatorname{Var}}

\newcommand{\diag}{\operatorname{diag}}

\newcommand{\CodeRepository}{\url{https://github.com/masahikoji/SIGA}}

\begin{document}
\setstretch{1.00}
\setlength{\emergencystretch}{2em}

\title{Fast Power Evaluation under Biased-Coin Minimization:
Sampling and Randomization Calibration}\author[1]{Masahiro Kojima}
\affil[1]{Department of Data Science for Business Innovation, Chuo University, 1-13-27 Kasuga, Bunkyo-ku, Tokyo 112-8551, Japan. E-mail: mkojima263@g.chuo-u.ac.jp}
\date{}
\maketitle

\noindent\textbf{Running title:} Fast Power Evaluation under Biased-Coin Minimization\\
\textbf{Corresponding author:} Masahiro Kojima, Department of Data Science for Business Innovation, Chuo University, 1-13-27 Kasuga, Bunkyo-ku, Tokyo 112-8551, Japan; E-mail: mkojima263@g.chuo-u.ac.jp.

\begin{abstract}
Design-stage power and sample-size evaluation under biased-coin minimization can be computationally intensive when a prespecified randomization test is reproduced within every simulated trial. We develop a reusable stratum-imbalance Gaussian approximation (SIGA) framework by exactly decomposing a fixed-score statistic into joint-stratum imbalance and orthogonal within-stratum components. Under explicit allocation-copy limit conditions for the same absolute-imbalance rule, the sampling-calibrated procedure, SIGA-S, consistently estimates the repeated-sampling variance at a marginal mean- or risk-difference boundary. At a nonsharp boundary, the conditional variance of a fixed-score randomization test can differ because the score contains the observed allocation path. To characterize this distinction, we express the first-order variance gap as a quadratic form involving pair-path covariance and introduce the randomization-calibrated procedure, SIGA-R, based on a reusable paired allocation-only calibration to approximate the conditional reference distribution. Separate comprehensive benchmarks showed close agreement between each SIGA procedure and the corresponding reference randomization test. A trial-inspired simulation based on published aggregate planning characteristics likewise produced similar power for SIGA-S, SIGA-R and the reference randomization test, while both reusable calibration procedures substantially reduced computation relative to nested rerandomization.
\end{abstract}

\section{Introduction}
Pocock--Simon minimization is widely used in clinical trials to balance treatment allocation across several prognostic factors, particularly when stratified permuted-block randomization based on all combinations of factor levels would create many strata relative to the total sample size, leaving some strata sparsely populated and some blocks incomplete \citep{PocockSimon1975,YeEtAl2022}. Regulatory guidance emphasizes prespecification of analyses that account for covariate-adaptive allocation, and randomization-based analyses have been discussed as an option when conventional procedures may not control type I error \citep{EMA2015Covariates,FDA2023Covariates,NMPA2022Randomization}.

Under minimization, treatment assignments are generated sequentially, with the allocation probability for a newly enrolled participant depending on the current marginal imbalances in prespecified baseline factors. Unrestricted permutation of the observed treatment labels therefore does not reproduce the original allocation law. One way to construct a randomization test that respects this law is to condition on the observed enrollment order and factor sequence and regenerate treatment paths from the same prespecified algorithm \citep{GailEtAl1988,ParhatEtAl2014,ProschanDodd2019,CoartEtAl2023}. Although such rerandomization is computationally feasible for the final analysis of a single trial, its use in design-stage power evaluation requires a nested Monte Carlo calculation. Specifically, a large number of treatment paths must be regenerated within each trial generated in the outer simulation loop to calculate its randomization-test $p$-value. Repeating this calculation across many simulated trials, candidate sample sizes, treatment effects, outcome models, analysis strategies and null boundaries can impose a substantial computational burden.

Inference following covariate-adaptive randomization has been studied under linear, generalized-linear and model-robust frameworks \citep{ShaoEtAl2010,ShaoYu2013,MaEtAl2015,YeEtAl2022}. For Pocock--Simon-type procedures, overall and marginal imbalances may be strongly controlled, whereas treatment imbalances within the joint strata formed by all minimization factors may fluctuate on the square-root scale of the total sample size \citep{HuZhang2020}. The limiting covariance of these joint-stratum imbalances is therefore relevant to treatment-effect statistics and can be estimated from the allocation mechanism \citep{ZhaoEtAl2024,KuznetsovaEtAlInPress}; related work has used this covariance structure to study test validity, including for time-to-event endpoints \citep{JohnsonEtAl2024}. These results suggest the possibility that the repeated-sampling law of a treatment-score statistic could be characterized through a design-specific imbalance component and an outcome-score component, thereby offering a potential route toward reducing, or ultimately avoiding, repeated simulation of the full randomization distribution.

Reducing the computational burden alone is not sufficient; the inferential distribution to be approximated must also be specified. We use as our reference procedure a design-based rerandomization test that holds an outcome-based score fixed and regenerates treatment assignments according to the original randomization mechanism \citep{GailEtAl1988,ParhatEtAl2014,ProschanDodd2019}. Under minimization, this entails keeping the observed enrollment order and factor sequence fixed and reapplying the planned minimization algorithm \citep{CoartEtAl2023}. For participant $i$, let $A_i\in\{0,1\}$ denote the treatment assignment and let $Y_i(a)$ denote the potential outcome under treatment $a\in\{0,1\}$. For a null boundary $b$, the score is constructed by subtracting $bA_i$ from the observed outcome and then centering or residualizing the resulting vector. If the constant-additive sharp null
\[
Y_i(1)-Y_i(0)=b
\]
is compatible with the outcome support, the boundary-adjusted outcomes are invariant to treatment assignment, and the resulting Fisher randomization test is finite-sample exact \citep{WuDing2021}. Such compatibility holds for unrestricted continuous outcomes and, for binary outcomes, at $b=0$; a nonzero binary risk-difference boundary generally does not admit a constant-additive sharp-null interpretation within the binary support. Throughout this article, we call this fixed-score procedure the reference randomization test, whether or not finite-sample exactness under a compatible sharp null is available.

The superiority, non-inferiority and equivalence hypotheses considered here instead concern a weak null on a marginal mean or risk difference and allow treatment effects to vary across participants. At such a nonsharp boundary, the boundary-adjusted score contains the observed allocation path, and the repeated-sampling distribution of the observed statistic need not agree with its conditional randomization distribution. Analogous distinctions motivate studentization and covariance adjustment in general permutation and Fisher-randomization theory \citep{Janssen1997,WuDing2021,ZhaoDing2021}. Under minimization, the discrepancy has an additional design-specific structure because the observed allocation path embedded in the score is combined with each regenerated path. Thus, computational acceleration and inferential target selection cannot be separated: an efficient approximation must be calibrated either to the repeated-sampling distribution of the observed statistic or to the conditional randomization distribution of the prespecified test.

In this paper, we develop a reusable stratum-imbalance Gaussian approximation (SIGA) framework for design-stage power and sample-size evaluation under Pocock--Simon minimization and propose SIGA-S as its primary sampling-calibrated procedure. Our contributions are threefold. First, we derive an exact orthogonal projection of a statistic formed from treatment assignments and a fixed score vector into a joint-stratum imbalance component and an orthogonal within-stratum component. This decomposition leads to a reusable design-level covariance calibration based only on allocation simulations and adapted to the realized joint-stratum counts. We show that the resulting SIGA-S variance estimator consistently estimates the repeated-sampling variance and yields asymptotically valid inference at a marginal mean- or risk-difference boundary.

Second, we formalize the dependence created by holding the score fixed at a nonsharp boundary. Because the fixed score contains the observed allocation path, evaluating that score along a regenerated path produces terms involving both the observed and regenerated paths. We call this observed--regenerated pairing the pair-path structure. We derive a design-specific identity for the resulting difference between the repeated-sampling and conditional randomization variances. This identity separates a genuine difference between the two inferential targets from error introduced by the Gaussian approximation, characterizes when the reference randomization test is asymptotically valid under the weak null and, when it is not, determines whether its first-order size distortion is conservative or anti-conservative.

Third, we make this pair-path structure estimable through a reusable allocation-only calibration. Conditional on the factor sequence, the observed path and a regenerated path are independent draws from the same allocation law. Their conditional joint law can therefore be reproduced by two independently generated paths under a common factor sequence. We use such simulated path pairs to estimate the pair-path covariance. This calibration yields SIGA-R for settings in which the rejection probability of the reference randomization test is the design target. When that test is prespecified as the final analysis, SIGA-R replaces the inner rerandomization loop within each simulated trial and provides a first-order approximation to its conditional randomization distribution. The same calibration also yields a variance ratio that quantifies the discrepancy between the randomization and repeated-sampling targets and indicates the resulting direction of weak-null size distortion.

SIGA-S and SIGA-R therefore serve different inferential targets. SIGA-S is the primary procedure for average-effect inference and the corresponding power and sample-size calculations, whereas SIGA-R is used when the design-stage objective is to approximate the rejection probability of the prespecified reference randomization test. They are not competing methods for a common inferential target. SIGA-R makes any discrepancy between the sampling and randomization targets explicit, but it does not correct weak-null size distortion of the reference test. Both design-level calibrations can be reused across outer simulation replicates, treatment effects, null boundaries, outcome models and prespecified baseline-only adjustment strategies.

Section~\ref{sec:methods} introduces the reference randomization test and the two SIGA procedures. Section~\ref{sec:theory} establishes their asymptotic properties, and Section~\ref{sec:algorithm} describes their use in design-stage power and sample-size evaluation. Numerical studies and the trial-inspired illustration are presented in Sections~\ref{sec:simulation} and \ref{sec:case-study}, followed by the discussion in Section~\ref{sec:discussion}. Full assumptions and proofs are provided in the Supplementary Material.

\section{Reference randomization test and the SIGA framework}\label{sec:methods}
This section first specifies the allocation rule and the reference randomization test. It then develops SIGA-S for repeated-sampling inference on the marginal treatment effect and SIGA-R for efficient design-stage reproduction of a prespecified fixed-score randomization test.

\subsection{Allocation rule and reference randomization test}\label{sec:fixed-score}
Consider a two-arm trial with total sample size $n$, in which $A_i\in\{0,1\}$ denotes the treatment assignment and $Z_i=2A_i-1\in\{-1,1\}$ its treatment sign. Let $X_{if}$ denote the level of factor $f\in\{1,\ldots,F\}$ for participant $i$, and let $S_i\in\{1,\ldots,J\}$ denote the joint stratum defined by all minimization factors. Write $x_f(s)$ for the level of factor $f$ in joint stratum $s$. To make the allocation mechanism explicit, we use a two-arm stochastic Pocock--Simon procedure with the range criterion \citep{PocockSimon1975,CoartEtAl2023}. For two treatments, this criterion sums the absolute marginal treatment imbalances under the two candidate assignments \citep{CoartEtAl2023}. As described by \citet{CoartEtAl2023}, we represent overall treatment balance by an additional one-level study factor and assign equal weight to all imbalance components. Define the overall imbalance immediately before participant $i$ is assigned by
\[
D_{i-1,0}=\sum_{j<i}Z_j
\]
and the level-specific marginal imbalance by
\[
D_{i-1,f\ell}
=
\sum_{j<i}Z_j\mathbf 1(X_{jf}=\ell).
\]
For a hypothetical treatment sign $z\in\{-1,1\}$ for participant $i$, define the imbalance criterion by
\begin{equation}\label{eq:min-score}
\mathcal I_i(z)
=
|D_{i-1,0}+z|
+
\sum_{f=1}^F|D_{i-1,f,X_{if}}+z|.
\end{equation}
Let $\operatorname{sgn}_0(x)=1$, $0$ or $-1$ according as $x>0$, $x=0$ or $x<0$, and define
\[
B_{i-1}(s)
=
\operatorname{sgn}_0(D_{i-1,0})
+
\sum_{f=1}^F
\operatorname{sgn}_0(D_{i-1,f,x_f(s)}).
\]
The elementary identity $|x+1|-|x-1|=2\operatorname{sgn}_0(x)$ gives
\begin{equation}\label{eq:absolute-sign-identity}
\mathcal I_i(1)-\mathcal I_i(-1)
=2B_{i-1}(S_i).
\end{equation}
Equation~\eqref{eq:absolute-sign-identity} shows that the range rule depends only on the signs of the active overall and marginal imbalances. This distinguishes it from squared-potential formulations, for which the difference between the two candidate scores also depends on the numerical magnitudes of the current imbalances. If $\mathcal I_i(1)<\mathcal I_i(-1)$, treatment 1 is assigned with probability $p_{\rm bc}>1/2$; if $\mathcal I_i(1)>\mathcal I_i(-1)$, treatment 1 is assigned with probability $1-p_{\rm bc}$; and a tie is resolved by assigning treatment 1 with probability $1/2$. The assumed factor distribution, the prespecified allocation rule, $p_{\rm bc}$ and the target sample size comprise the planned design $\mathcal D_n$.

Supplementary Appendix A.2 establishes several exact properties of this absolute-imbalance range rule and states the one-path and three-path allocation-limit conditions used in the asymptotic theory. The one-path condition concerns the observed allocation path. The three-path condition is used for the conditional-randomization theory and jointly treats the observed path and two paths regenerated independently conditional on the same factor sequence. The second regenerated path is an auxiliary conditionally independent replicate used to establish convergence of the conditional randomization law, as described in Supplementary Appendix C.1; it is not an additional path required by the routine SIGA-R calibration. Supplementary Appendix A.2 also shows that a suitable one-path finite-step drift condition would imply the required allocation limits for any fixed number of paths. The exact score projection and both allocation-only calibrations are finite-sample constructions and do not rely on these asymptotic conditions.

Let $\bm r=(r_1,\ldots,r_n)^\top$ be a centered score vector satisfying $\bm 1_n^\top\bm r=0$, and let $\bm Z=(Z_1,\ldots,Z_n)^\top$ collect the treatment signs. For $a\in\{0,1\}$, let $n_a=\sum_{i=1}^n\mathbf 1(A_i=a)$ and, when $n_a>0$, let $\bar r_a=n_a^{-1} \sum_{i=1}^n\mathbf 1(A_i=a)r_i$. Define
\begin{equation}\label{eq:T-score}
T_n(\bm r)
=
\sum_{i=1}^n\left(A_i-\frac12\right)r_i
=
\frac12\bm Z^\top\bm r
=
\frac{n_1n_0}{n}(\bar r_1-\bar r_0),
\end{equation}
where the final equality is used on the event $n_0n_1>0$.

Let $Y_i(a)$ be the potential outcome under assignment $a$, let $Y_i=A_iY_i(1)+(1-A_i)Y_i(0)$, and let $\Delta=\E[Y_i(1)-Y_i(0)]$ be the marginal mean difference; for a binary outcome, this is the marginal risk difference. For a prespecified null boundary $b$ on the scale of the marginal treatment effect $\Delta$, define the boundary-adjusted outcome $W_i(b)=Y_i-bA_i$. The unadjusted score is $\bm r(b)=\bm W(b)-\bar W(b)\bm 1_n$. For a prespecified baseline-only adjustment, let $\bm C$ contain an intercept and a fixed number of baseline functions. On the event that $\bm C^\top\bm C$ is nonsingular, define $\bm r(b)=\bm M_C\bm W(b)$, $\bm M_C=\bm I_n-\bm C(\bm C^\top\bm C)^{-1}\bm C^\top$. The adjustment model contains no treatment indicator and is fitted once; the resulting residual score is then held fixed over all regenerated treatment paths. Under the participant-level conditions in Supplementary Appendix B.1, $n^{-1}\bm C^\top\bm C$ converges in probability to the positive-definite matrix $\bm\Sigma_C$. Hence, $\bm C^\top\bm C$ is nonsingular with probability tending to one; a Moore--Penrose inverse may be used in finite samples if needed.

For continuous or binary outcomes, viewed on their numerical scale, the boundary-adjusted observation has the exact algebraic representation
\begin{equation}\label{eq:Qdelta-identity}
W_i(b)
=
Q_i(b)+\frac12Z_i\delta_i(b),
\qquad
Q_i(b)
=
\frac{Y_i(1)+Y_i(0)-b}{2},
\qquad
\delta_i(b)
=
Y_i(1)-Y_i(0)-b.
\end{equation}
At the boundary $\Delta=b$, $\E[\delta_i(b)]=0$. Under a compatible constant-additive sharp null, $Y_i(1)-Y_i(0)=b$ for every participant and hence $\delta_i(b)=0$. As noted above, compatibility with binary potential outcomes generally requires $b=0$. More generally, if $d_s(b)=\E[\delta_i(b)\mid S_i=s]=0$ $(s=1,\ldots,J)$, the average-effect boundary holds within every joint stratum even though individual effects may be heterogeneous.

The reference randomization test considered here follows the design-based rerandomization principle of holding an outcome-based score fixed while regenerating treatment assignments under the original randomization mechanism \citep{GailEtAl1988,ParhatEtAl2014,ProschanDodd2019}. In our boundary-adjusted implementation, $\bm r(b)$ is constructed from the observed data and then held fixed. For minimization, we condition on the observed ordered factor sequence and regenerate treatment paths by reapplying the prespecified minimization algorithm \citep{CoartEtAl2023}. More precisely, it generates independent paths $\bm Z_1^*,\ldots,\bm Z_B^*$ and calculates
\[
T_{n,j}^*\{\bm r(b)\}
=
\frac12(\bm Z_j^*)^\top\bm r(b).
\]
Because the same score vector $\bm r(b)$ enters the observed statistic and every regenerated statistic, replacing $\bm r(b)$ by $c\bm r(b)$ for any $c>0$ multiplies all these statistics by $c$ and therefore leaves the corresponding one- and two-sided randomization-test $p$-values unchanged. For a two-sided test, the inclusive plus-one Monte Carlo $p$-value \citep{PhipsonSmyth2010} is
\[
\widehat p_{\rm RT}
=
\frac{
1+
\sum_{j=1}^B
\mathbf 1\left[
|T_{n,j}^*\{\bm r(b)\}|
\ge
|T_n\{\bm r(b)\}|
\right]
}{
B+1
}.
\]
One-sided tests use the corresponding tail. Equivalence with limits $L<U$ is assessed by the intersection--union procedure based on the upper-tail test at $b=L$ and the lower-tail test at $b=U$. Exactness under a compatible sharp null does not imply validity for a weak null on a marginal mean or risk difference. At a nonsharp boundary---including a nonzero binary risk-difference boundary---the repeated-sampling law of the observed statistic and the conditional randomization law are therefore treated separately below.

\subsection{Sampling-calibrated SIGA}\label{sec:sampling-siga}
Let $\bm H$ be the $n\times J$ joint-stratum indicator matrix and let $\bm N=\bm H^\top\bm H=\diag(N_1,\ldots,N_J)$, where $N_s$ is the number of participants in joint stratum $s$. Zero-count strata are handled by the Moore--Penrose inverse $\bm N^+$. Define
\[
\bm P_H
=
\bm H\bm N^+\bm H^\top,
\qquad
\bm D
=
\bm H^\top\bm Z,
\qquad
\bar{\bm r}
=
\bm N^+\bm H^\top\bm r,
\qquad
\bm e
=
(\bm I_n-\bm P_H)\bm r.
\]
Then $\bm H^\top\bm e=\bm0$ and the exact decomposition
\[
T_n(\bm r)
=
\frac12\bar{\bm r}^{\top}\bm D
+
\frac12\bm Z^\top\bm e
\]
holds. A proof is given in Supplementary Appendix A.1. The first term is driven by treatment imbalance within the joint strata and the second by within-stratum score contrasts.

For each candidate design and sample size, generate $B_0$ allocation-only replicates. In replicate $q$, let $\bm N^{(q)}$ and $\bm D^{(q)}$ denote the corresponding joint-stratum count matrix and treatment-sign imbalance vector. Define
\[
U_s^{(q)}
=
\begin{cases}
D_s^{(q)}/\sqrt{N_s^{(q)}},&N_s^{(q)}>0,\\
0,&N_s^{(q)}=0,
\end{cases}
\]
and let $\widehat{\bm\Gamma}_n$ be the sample covariance of the vectors $\bm U^{(q)}$. For a realized count matrix $\bm N$, set $\widetilde{\bm\Omega}_n(\bm N)=\bm N^{1/2}\widehat{\bm\Gamma}_n\bm N^{1/2}$. The covariance completion is $\widehat{\bm\Sigma}_{Z,n}(\bm N)=\bm H\bm N^+\widetilde{\bm\Omega}_n(\bm N)\bm N^+\bm H^\top+\widehat\kappa_n(\bm I_n-\bm P_H)$, where, when $n>J_+$,
\[
\widehat\kappa_n^{\rm raw}
=
\frac{
n-
\operatorname{tr}
\{\bm N^+\widetilde{\bm\Omega}_n(\bm N)\}
}{
n-J_+
},
\qquad
\widehat\kappa_n
=
\max(\widehat\kappa_n^{\rm raw},0),
\]
where $J_+=\sum_{s=1}^J\mathbf 1(N_s>0)$ is the number of distinct combinations of minimization-factor levels represented by at least one participant in the realized sample. If $n=J_+$, every represented joint stratum contains exactly one participant, so there is no nontrivial within-stratum residual subspace; set $\widehat\kappa_n=0$, and the second term vanishes. The first term reproduces the calibrated covariance of $\bm D$ on the nonempty joint strata, whereas the second assigns a common variance to the orthogonal within-stratum subspace. When $\widehat\kappa_n^{\rm raw}\ge0$, the coefficient matches the total trace to $n$; the positive-part safeguard preserves positive semidefiniteness otherwise. These properties and the trace-matching derivation are given in Supplementary Appendix A.1.

The resulting variance estimator is
\begin{equation}\label{eq:VS-main}
\widehat V_{{\rm S},n}(\bm r)
=
\frac14
\bm r^\top
\widehat{\bm\Sigma}_{Z,n}(\bm N)
\bm r
=
\frac14
\left\{
\bar{\bm r}^\top
\widetilde{\bm\Omega}_n(\bm N)
\bar{\bm r}
+
\widehat\kappa_n
\bm e^\top\bm e
\right\}.
\end{equation}
The normal-approximation procedure based on \eqref{eq:VS-main} is denoted SIGA-S. As shown in Section~\ref{sec:theory}, $\widehat V_{{\rm S},n}\{\bm r(b)\}$ consistently estimates the repeated-sampling variance of the observed statistic at an average-effect boundary. This boundary result establishes null validity. Away from the boundary, design-stage power is the rejection probability of the same boundary-calibrated test and does not require the variance estimator to estimate the centered sampling variance under the alternative. Its standardized statistic is
\[
Z_{{\rm S},n}(b)
=
\frac{
T_n\{\bm r(b)\}
}{
\widehat V_{{\rm S},n}\{\bm r(b)\}^{1/2}
}.
\]
At a nonsharp boundary, however, the repeated-sampling variance need not equal the conditional randomization variance of the reference test. The pair-path calibration developed next addresses this distinct variance target.

\subsection{Pair-path calibration and randomization-targeted SIGA-R}\label{sec:pair-calibration}
SIGA-R is intended for a specific design-stage task: calculating power or sample size for a final analysis that has already been prespecified as the fixed-score reference randomization test. Directly reproducing that analysis requires an inner rerandomization loop within every outer simulated trial. SIGA-R replaces the repeated inner loops by a reusable allocation-only calibration, while retaining the conditional randomization variance targeted by the reference test. It is therefore useful when the design-stage calculation must correspond to the prespecified randomization analysis, rather than when the primary target is repeated-sampling inference on the marginal treatment effect.

The distinction can matter at a nonsharp boundary when the stratum-specific deviations $d_s(b)$ are not all zero. A concrete example is a binary non-inferiority or equivalence design with a nonzero risk-difference margin and treatment effects that vary across the minimization strata. By \eqref{eq:Qdelta-identity}, the score held fixed by the reference randomization test then contains the observed allocation path through $Z_i\delta_i(b)/2$. A statistic evaluated on a regenerated path therefore contains products of the regenerated and observed treatment signs. Consequently, the stratum-mean heterogeneity contribution to the conditional randomization variance is governed by the covariance of a pair-path profile process, whereas the corresponding contribution to the repeated-sampling variance is governed by the joint-stratum probability matrix. Their difference produces the variance gap derived below. In such a setting, SIGA-S and the reference randomization test can use different first-order reference variances; SIGA-R estimates the variance needed to reproduce the latter. The practical consequence is that a design based on SIGA-S can overstate or understate the power of the prespecified randomization test and can therefore select a different sample size. This gap reflects a difference between inferential targets, rather than an approximation error of SIGA-S.

For a candidate design and sample size, estimate the pair-path covariance using $B_\Psi$ independent allocation-only calibration replicates. In replicate $q$, generate one factor sequence and apply the minimization algorithm twice using independent randomization numbers, yielding two complete allocation paths $\bm Z^{(q,1)}$ and $\bm Z^{(q,2)}$ conditional on the same sequence. Each path contains assignments to both treatment groups; the two paths do not correspond to the two treatment arms. Two paths are the minimum required to construct the pair-product process whose covariance defines the SIGA-R correction. More than two paths can provide additional but mutually dependent pair products, so the gain is not proportional to the number of paths. The routine implementation therefore uses two paths per replicate and controls Monte Carlo precision through the number $B_\Psi$ of independent calibration replicates. Define
\begin{equation}\label{eq:G-pair-main}
\bm G_n^{(q,12)}
=
\frac1{\sqrt n}
\sum_{i=1}^n
\bm h(S_i^{(q)})
Z_i^{(q,1)}
Z_i^{(q,2)},
\end{equation}
where $\bm h(s)$ is the $J$-vector indicating joint stratum $s$. Let $\widehat{\bm\Psi}_n$ be the sample covariance of the vectors in \eqref{eq:G-pair-main}. For an outer trial with joint-stratum count matrix $\bm N$, set $\widehat{\bm\Pi}_n=\bm N/n$. Thus $\widehat{\bm\Psi}_n$ estimates the design-level covariance of the normalized pair-path process, whereas $\widehat{\bm\Pi}_n$ consistently estimates the joint-stratum probability matrix. The third path appearing in the asymptotic theory is not part of this routine calibration: it supplies the second regenerated replicate required by the conditional-copy argument and permits cross-path covariance diagnostics in Supplementary Appendices A.2 and C.1.

Let $\widehat{\bm d}(b)$ be a consistent estimate of the vector of stratum-specific boundary deviations $\bm d(b)$. In design-stage simulation, the natural choice is the value implied by the prespecified outcome-generating model. A data-based consistent construction is given in Supplementary Appendix C.2 to complete the asymptotic formulation, but the principal application of SIGA-R is design-stage reproduction of the reference randomization test. Define the estimated pair-path variance-gap term by
\begin{equation}\label{eq:pair-gap-main}
\widehat C_n(b)
=
\frac n{16}
\widehat{\bm d}(b)^\top
\{\widehat{\bm\Psi}_n-\widehat{\bm\Pi}_n\}
\widehat{\bm d}(b),
\end{equation}
and define the unsafeguarded randomization-calibrated variance by
\begin{equation}\label{eq:VRraw-main}
\widehat V_{{\rm R},n}^{\rm raw}(b)
=
\widehat V_{{\rm S},n}\{\bm r(b)\}
+
\widehat C_n(b).
\end{equation}
Because $\widehat{\bm\Psi}_n-\widehat{\bm\Pi}_n$ need not be positive semidefinite, the variance-gap term can have either sign. For numerical protection, set
\begin{equation}\label{eq:VR-main}
\widehat V_{{\rm R},n}(b)
=
\max\left\{
\widehat V_{{\rm R},n}^{\rm raw}(b),
\epsilon_n
\widehat V_{{\rm S},n}\{\bm r(b)\}
\right\},
\end{equation}
where $\epsilon_n>0$ and $\epsilon_n\to0$. The safeguard is asymptotically inactive when the limiting randomization variance is positive.

The normal-approximation procedure based on \eqref{eq:VR-main} is denoted SIGA-R, with standardized statistic
\[
Z_{{\rm R},n}(b)
=
\frac{
T_n\{\bm r(b)\}
}{
\widehat V_{{\rm R},n}(b)^{1/2}
}.
\]
SIGA-S and SIGA-R use the same observed statistic $T_n\{\bm r(b)\}$ and differ only in the variance used to calibrate its reference distribution. The practical choice is therefore determined by the prespecified final analysis. Use SIGA-S when the intended analysis is inference on the marginal average effect using the sampling-calibrated normal approximation; at the tested boundary, its variance estimator targets the repeated-sampling variance of the observed statistic. Use SIGA-R when the intended final analysis is the fixed-score reference randomization test and the design-stage power or sample size must reproduce that test's rejection probability. In the nonsharp and stratum-heterogeneous settings described above, using SIGA-S for that purpose can calibrate to a different first-order variance from the one used by the randomization test, whereas SIGA-R applies the pair-path correction to approximate the test's conditional reference distribution, including under fixed alternatives. If the two variance targets align, SIGA-R offers no inferential advantage and reduces to the same first-order calibration as SIGA-S. Thus SIGA-R is not a generally preferable successor to SIGA-S; it is a computational surrogate for a prespecified randomization-based analysis.

The estimated variance ratio
\begin{equation}\label{eq:rho-main}
\widehat\rho_n(b)
=
\frac{
\widehat V_{{\rm R},n}^{\rm raw}(b)
}{
\widehat V_{{\rm S},n}\{\bm r(b)\}
}
=
1+
\frac{
\widehat C_n(b)
}{
\widehat V_{{\rm S},n}\{\bm r(b)\}
}
\end{equation}
is the primary diagnostic of the pair-path correction. The unsafeguarded variance is used in \eqref{eq:rho-main} so that the numerical protection in \eqref{eq:VR-main} does not conceal an unstable pair-path estimate. Values close to one indicate that the correction is practically negligible, whereas material departures from one indicate a potentially meaningful change in the reference calibration. At a weak-null boundary, the population ratio $v_{\rm R}(b)/v_{\rm S}(b)$ compares the conditional-randomization and repeated-sampling variance targets and determines the direction and magnitude of the reference test's size distortion through Corollary~\ref{cor:main-alignment}. Away from the boundary, $\widehat\rho_n(b)$ remains a useful design-stage measure of the pair-path adjustment, but its denominator is the SIGA-S score quadratic rather than an asserted estimator of the centered sampling variance under the fixed alternative. There is no universal cutoff for a material departure; its practical relevance should be assessed through the resulting change in rejection probability or selected sample size. If $\widehat V_{{\rm R},n}^{\rm raw}(b)\le0$ or the safeguard is active, the result should be treated as a numerical warning and checked using a larger paired calibration and direct rerandomization.

If $\bm d(b)=\bm0$, including a compatible constant-additive sharp boundary and a boundary that holds on average within every joint stratum, the population pair-path variance gap is zero. When this condition is imposed in a design-stage model, setting $\widehat{\bm d}(b)=\bm0$ makes SIGA-S and SIGA-R identical. The condition $\bm d(b)=\bm0$ is sufficient but not necessary; the more general first-order alignment condition is given in Corollary~\ref{cor:main-alignment}.

SIGA-R reproduces the conditional reference distribution of the reference randomization test, including any weak-null size distortion of that test; it does not correct such distortion. Accordingly, it should be selected because the randomization test is the prespecified final analysis, not merely because SIGA-R is an extension of SIGA-S or because it yields a different power estimate. SIGA-R is a design-stage surrogate for that prespecified analysis, whereas the randomization test itself remains the reference procedure for the final analysis.

\section{Asymptotic theory}\label{sec:theory}
We first state conditions on the allocation process independently of the outcome model. For the conditional-randomization argument, the observed allocation path is indexed by 0, and two additional regenerated allocation paths are indexed by 1 and 2. Conditional on a common profile sequence, all three paths are generated by the same allocation rule using mutually independent randomization variables. Path 1 represents a generic draw from the conditional randomization distribution, whereas path 2 is an independent duplicate used in the conditional-copy argument establishing convergence of that distribution. Accordingly, the one-path limit for the observed path governs the repeated-sampling distribution, the two-path marginal for paths 0 and 1 determines the pair-path covariance, and the joint three-path limit for paths 0, 1 and 2 supports the conditional convergence result.

The number of minimization factors and their numbers of levels are fixed, and participant profiles are independent across participants with positive joint-profile probabilities. For SIGA-S, Supplementary Appendix A.2 assumes the joint Gaussian limit and second-moment convergence of the profile-count fluctuation and the joint-stratum imbalance for one allocation path. For SIGA-R, the same subsection assumes the corresponding three-path joint limit, including the pair-path profile processes and tightness of the triple-product process. These are the minimal allocation-level conditions used in the respective proofs. Label-flip symmetry in Supplementary Appendix A.2 identifies the zero cross-covariance blocks.

Supplementary Appendix A.2 proves several exact properties of the absolute-imbalance range-rule chain in \eqref{eq:min-score} and shows that, if a one-path finite-step geometric drift condition holds, the required allocation limits follow for any fixed number of paths. It also shows that the natural one-step $\ell_1$ Lyapunov function has positive drift on an unbounded family of reachable states. These results do not, however, establish the required one-path stability condition for the full class of positive joint-profile distributions considered here. The allocation-limit results below are therefore stated under the explicit one-path and three-path conditions in Supplementary Appendix A.2. Existing results for squared-potential imbalance criteria cannot fill this gap because those criteria can prefer a different assignment from the range rule.

For the adjusted score, let $\bm c_i$ be the baseline vector and define $\bm\beta_b=\{\E[\bm c_i\bm c_i^\top]\}^{-1}\E[\bm c_iQ_i(b)]$ and $q_i(b)=Q_i(b)-\bm c_i^\top\bm\beta_b$, together with $m_s(b)=\E[q_i(b)\mid S_i=s]$ and $\nu_i^Q(b)=q_i(b)-m_{S_i}(b)$. For the unadjusted score, $\bm c_i$ contains only the intercept. Let $\bm m(b)$ and $\bm d(b)$ collect $m_s(b)$ and $d_s(b)$, respectively, let $\bm\Pi=\diag(\pi_1,\ldots,\pi_J)$, let $\bm\Sigma_D$ be the limiting covariance of $\bm D/\sqrt n$, and let $\bm\Psi_D$ be the limiting covariance of the pair-path profile process.

The following results provide formal guarantees for the two proposed procedures. Theorem~\ref{thm:main-sampling} shows that SIGA-S consistently estimates the repeated-sampling variance and yields asymptotically valid inference at an average-effect boundary. Theorem~\ref{thm:main-randomization} shows that SIGA-R consistently estimates the conditional randomization variance and is asymptotically equivalent to the ideal version of the reference randomization test. Corollary~\ref{cor:main-alignment} characterizes when the two variance targets coincide and quantifies the resulting size distortion and local power when they do not. Supplementary Appendix B.1 develops the participant-level expansions, including the baseline-only adjustment remainder, and Supplementary Appendix B.2 proves Theorem~\ref{thm:main-sampling}. Supplementary Appendices C.1 and C.2 prove Theorem~\ref{thm:main-randomization}, whereas Supplementary Appendix C.3 proves Corollary~\ref{cor:main-alignment} and gives the intersection--union extension for equivalence.

\begin{theorem}\label{thm:main-sampling}
Suppose the one-path allocation-limit condition in Supplementary Appendix A.2, the participant-level conditions in Supplementary Appendix B.1 and the one-path calibration conditions in Supplementary Appendix A.3 hold, and that $\Delta=b$. Then, for either the centered unadjusted score or the prespecified baseline-only adjusted score,
\[
\frac{T_n(\bm r(b))}{\sqrt n}
\xrightarrow{d}
\mathcal N(0,v_{\rm S}(b)),
\qquad
\frac{
\widehat V_{{\rm S},n}(\bm r(b))
}{n}
\xrightarrow{p}
v_{\rm S}(b),
\]
where
\begin{equation}\label{eq:vS-main}
v_{\rm S}(b)
=
\frac14\left[
\bm m(b)^\top\bm\Sigma_D\bm m(b)
+
\E[\nu_i^Q(b)^2]
+
\frac14\E[\delta_i(b)^2]
\right].
\end{equation}
If $v_{\rm S}(b)>0$, then
\[
Z_{{\rm S},n}(b)
\xrightarrow{d}
\mathcal N(0,1).
\]
Hence, the SIGA-S $p$-value is asymptotically valid at the average-effect boundary without requiring alignment of the sampling and randomization variances.
\end{theorem}

\begin{theorem}\label{thm:main-randomization}
Fix a boundary value $b$, not necessarily equal to the data-generating marginal effect. Suppose the three-path allocation-limit condition in Supplementary Appendix A.2 and the participant-level conditions in Supplementary Appendix B.1 hold for the exact absolute-range rule. The three-path process consists of the observed allocation path $\bm Z^{(0)}$ and two additional allocation paths $\bm Z^{(1)}$ and $\bm Z^{(2)}$. Conditional on the common profile sequence, the three paths follow the prespecified allocation rule and are mutually independent, with $\bm Z^{(0)}$ denoting the path that generated the observed data.

Let $\mathcal O_n$ denote the $\sigma$-field generated by the observed profile sequence, baseline variables, observed outcome vector and observed allocation path. Since the resulting score $\bm r(b)$ is $\mathcal O_n$-measurable, the conditional law of the statistic based on a regenerated allocation path satisfies
\[
d_{\mathrm{BL}}\left(
\mathcal L\left(
\left.
\frac{
T_n^*(\bm r(b))
}{
\sqrt n
}
\right|
\mathcal O_n
\right),
\mathcal N(0,v_{\rm R}(b))
\right)
\xrightarrow{p}
0,
\]
where $d_{\mathrm{BL}}$ denotes the bounded-Lipschitz metric on the space of Borel probability measures on $\mathbb R$, and
\begin{equation}\label{eq:vR-main}
v_{\rm R}(b)
=
\frac14\left(
\bm m(b)^\top\bm\Sigma_D\bm m(b)
+
\E[\nu_i^Q(b)^2]
+
\frac14\E[\Var[\delta_i(b)\mid S_i]]
+
\frac14
\bm d(b)^\top
\bm\Psi_D
\bm d(b)
\right).
\end{equation}

If, in addition, the paired calibration conditions in Supplementary Appendix A.3 hold, $v_{\rm R}(b)>0$ and $\widehat{\bm d}(b)\xrightarrow{p}\bm d(b)$, then
\[
\frac{
\widehat V_{{\rm R},n}(b)
}{n}
\xrightarrow{p}
v_{\rm R}(b).
\]
Consequently, the $p$-values from the ideal version of the reference randomization test and SIGA-R are asymptotically equivalent for one- and two-sided tests, without requiring variance alignment. This conditional approximation applies both at a null boundary and under fixed alternatives used in design-stage power calculations. The corresponding Monte Carlo randomization-test $p$-value with the $+1$ correction is also asymptotically equivalent provided that the number of regenerated allocation paths satisfies $B=B_n\to\infty$.
\end{theorem}

Let $\Phi$ denote the standard normal distribution function and let $z_\gamma=\Phi^{-1}(\gamma)$.

\begin{corollary}\label{cor:main-alignment}
At $\Delta=b$,
\begin{equation}\label{eq:gap-main}
v_{\rm R}(b)-v_{\rm S}(b)
=
\frac1{16}
\bm d(b)^\top
(\bm\Psi_D-\bm\Pi)
\bm d(b).
\end{equation}
Thus, SIGA-S and SIGA-R coincide to first order, and the reference randomization test is asymptotically valid under the weak null, whenever
\[
\bm d(b)^\top
\bm\Psi_D
\bm d(b)
=
\bm d(b)^\top
\bm\Pi
\bm d(b).
\]
This condition holds automatically under any compatible constant-additive sharp null and whenever $d_s(b)=0$ for every joint stratum. For an upper-tail test of level $\alpha$, if alignment fails, the limiting rejection probability at the boundary is
\begin{equation}\label{eq:size-distortion-main}
1-\Phi\left(
z_{1-\alpha}
\sqrt{
v_{\rm R}(b)/v_{\rm S}(b)
}
\right).
\end{equation}
For an absolute-value two-sided test, replace $z_{1-\alpha}$ by $z_{1-\alpha/2}$ and multiply the upper-tail probability by two. Under the triangular-array local alternative in Supplementary Appendix B.2, satisfying $\Delta_n=b+h/\sqrt n$, the limiting upper-tail rejection probability is
\begin{equation}\label{eq:local-power-main}
1-\Phi\left(
\frac{
z_{1-\alpha}\sqrt{v_{\rm R}(b)}-h/4
}{
\sqrt{v_{\rm S}(b)}
}
\right).
\end{equation}
The corresponding SIGA-S limit is obtained by replacing $v_{\rm R}(b)$ in the critical value with $v_{\rm S}(b)$.
\end{corollary}

Taken together, these results separate two guarantees that are distinct under a nonsharp boundary. Under the stated conditions, SIGA-S provides asymptotically valid inference for the marginal average effect, whereas SIGA-R provides a first-order approximation to the conditional randomization distribution of the reference test. The variance-gap formula determines whether the reference randomization test itself is asymptotically valid under the weak null and, otherwise, whether its first-order size distortion is conservative or anti-conservative. For design-stage calculations, the asymptotic equivalence of SIGA-R justifies replacing nested rerandomization with the reusable allocation-only calibration based on paired paths. Weak-null validity of the reference test, however, additionally depends on the variance-alignment condition in Corollary~\ref{cor:main-alignment}.

\section{Design-stage power and sample-size evaluation}
\label{sec:algorithm}

The procedure used for sample-size selection is determined by the prespecified final analysis whose repeated-sampling rejection probability is to be targeted. SIGA-S is used for power and sample-size evaluation of the boundary-valid marginal average-effect test. SIGA-R is used when the planned final analysis is the reference fixed-score randomization test, because it approximates that test's conditional randomization distribution within each outer trial, including under fixed alternatives.

Before a randomization test is used as the target for sample-size selection, its rejection probability should also be examined at each relevant weak-null boundary. A value of $\widehat\rho_n(b)$ materially different from one indicates that the reference test may not attain its nominal weak-null level. SIGA-R can still reproduce that test, but it does not correct its size distortion.

For each candidate total sample size $n$, the allocation-only calibration estimates $\widehat{\bm\Gamma}_n$ and, when the lattice-normal mixture refinement is used, the treatment-count probabilities. A paired calibration additionally estimates $\widehat{\bm\Psi}_n$. These quantities are saved and reused across outer trials generated under the same minimization design and sample size.

For the SIGA-S analysis of unadjusted binary superiority at the boundary $b=0$, we use a lattice-normal mixture refinement of the tail probability. The refinement assigns continuity-corrected normal masses to the feasible lattice points conditional on each possible treatment-group size, normalizes over the feasible support and averages with respect to the treatment-count distribution retained from the allocation-only calibration. It uses the same observed statistic and the same SIGA-S variance estimate as the ordinary Gaussian calculation and modifies only the tail-probability calculation. Baseline-adjusted binary superiority, non-inferiority and equivalence use the ordinary Gaussian SIGA-S tail probabilities. SIGA-R likewise uses the ordinary Gaussian tail probabilities based on its randomization-calibrated variance. The lattice-normal mixture formulas are given in Supplementary Appendix D.

In outer trial $\ell$, the score $\bm r_\ell(b)$ is constructed at the relevant boundary. SIGA-S uses \eqref{eq:VS-main}; to approximate the reference randomization test at a nonsharp boundary, SIGA-R additionally uses the model-based or consistently estimated $\widehat{\bm d}_\ell(b)$ in \eqref{eq:VRraw-main}. Let $p_\ell(n)$ denote the selected SIGA $p$-value: the lattice-normal mixture $p$-value for unadjusted binary superiority analyzed with SIGA-S and the ordinary Gaussian $p$-value otherwise. Power is estimated by
\begin{equation}\label{eq:power-estimate}
\widehat{\operatorname{Power}}(n)
=
\frac1{B_{\rm out}}
\sum_{\ell=1}^{B_{\rm out}}
\mathbf 1\{p_\ell(n)\le\alpha\}.
\end{equation}
For equivalence, $p_\ell(n)$ is the maximum of the two one-sided $p$-values. The selected sample size is the smallest candidate satisfying the target power. Calibration Monte Carlo error can be assessed by repeating the allocation-only calibration using independent random-number streams.

\section{Simulation study}\label{sec:simulation}

The numerical studies evaluated the operating performance and computational efficiency of the proposed procedures. Two independent simulation studies were conducted using the same 56 scenario configurations. The first study evaluated SIGA-S as a boundary-valid repeated-sampling procedure by comparing its rejection probabilities with those of the reference randomization test. The second study evaluated SIGA-R as a computational approximation to the conditional reference distribution of the randomization test. Because SIGA-S and SIGA-R target different reference distributions, these comparisons serve distinct purposes and are not interpreted as a head-to-head ranking of the two procedures.

The 56 scenario configurations were obtained by crossing two outcome types, continuous and binary, with four minimization design--sample-size combinations and seven testing settings. The four design--sample-size combinations used either two independent binary minimization factors with 100 or 500 participants per group, or five independent binary minimization factors with 200 or 1000 participants per group. The seven testing settings covered superiority, non-inferiority and equivalence at null boundaries and under fixed alternatives. Each scenario was analyzed using both an unadjusted score and a baseline-adjusted score. Both simulation studies used the equal-weight absolute-imbalance range rule in \eqref{eq:min-score} with $p_{\rm bc}=0.80$; ties were resolved with probability $0.50$ for each treatment.

The first simulation study focused on SIGA-S. Its purpose was to evaluate SIGA-S as a boundary-valid repeated-sampling procedure and to determine how closely its rejection probabilities agreed with those of the reference randomization test under the prespecified simulation settings. A one-path allocation-only calibration with 100,000 replicates was used to estimate $\widehat{\bm\Gamma}_n$ and, when needed, the treatment-count distribution for the lattice-normal mixture. SIGA-S and the reference randomization test were then evaluated on the same $B_{\rm out}=100{,}000$ outer trials within each scenario.

The second simulation study focused on SIGA-R. Its purpose was to evaluate SIGA-R as a computational approximation to the conditional reference distribution of the randomization test. This study was conducted as an independent repeat of the same 56 scenarios and used 100,000 three-path allocation-only calibration replicates to estimate $\widehat{\bm\Gamma}_n$, $\widehat{\bm\Psi}_n$ and the treatment-count distribution. The SIGA-R safeguard was set to $\epsilon_n=1/n$. Within this independent study, SIGA-R and the reference randomization test were evaluated on common outer trials. Because the two simulation studies used independent outer simulations, their reference-test estimates need not be numerically identical.

In both simulation studies, each reference randomization test used $B=4{,}999$ regenerated paths and the inclusive plus-one Monte Carlo $p$-value, with ties included in the exceedance count. Within an outer trial, one common set of regenerated paths was used for all required null boundaries and both score analyses. SIGA-R used the model-implied vector $\bm d(b)$ so that its comparison with the reference test evaluated approximation of the conditional randomization distribution without adding estimation error from the stratum-specific effects.

An additional targeted pair-path sensitivity analysis was conducted to assess settings in which the pair-path correction materially affects approximation of the reference randomization test. Because this analysis was designed as a focused sensitivity assessment, its complete configuration and results are reported in Supplementary Appendix G. The analysis comprised 12 weak-null scenarios under six independent-factor allocation designs. Four designs used $p_{\rm bc}=0.80$, with two factors and 100 or 500 participants per group or five factors and 200 or 1000 participants per group; two additional two-factor designs used $p_{\rm bc}=0.95$ with 100 or 500 participants per group. Scenarios S01--S08 used the realistic continuous model or the common-log-odds binary model, whereas S09--S12 used deliberately strong maximum-ratio heterogeneity directions selected using an independent allocation-only calibration. Each design used 100,000 three-copy analysis-calibration replicates, the strong directions used a separate 200,000-replicate direction-selection calibration, each scenario used 100,000 outer trials and each reference randomization test used 4,999 regenerated paths. Complete scenario configurations are provided in Supplementary Table~3, and rejection probabilities are reported in Supplementary Table~4.

\subsection{Continuous-outcome study}\label{sec:continuous-sim}

The continuous-outcome study provided a regular setting with a homogeneous additive treatment effect across joint strata. It was not designed to make the pair-path variance discrepancy artificially large. Four design--sample-size combinations were considered. With two independent binary minimization factors, the factor prevalences were
\[
\Pbb(X_{i1}=1)=0.50,
\qquad
\Pbb(X_{i2}=1)=0.40,
\]
and the nominal sample sizes were 100 and 500 participants per group. With five independent binary minimization factors, the corresponding prevalences were
\[
(0.50,0.40,0.30,0.20,0.10),
\]
and the nominal sample sizes were 200 and 1000 participants per group.

The control potential outcome was generated as
\[
Y_i(0)
=
\sum_{j=1}^F
\beta_j\{X_{ij}-\Pbb(X_{ij}=1)\}
+
\xi\{X_{i1}X_{i2}-\E[X_{i1}X_{i2}]\}
+
\varepsilon_i,
\]
where $F$ is the number of minimization factors, $\varepsilon_i\sim\mathcal N(0,1)$ and $\xi=0.25$. For simplicity, the main-effect coefficient vector was set equal to the corresponding vector of factor prevalences. Thus,
\[
\bm\beta=(0.50,0.40)^\top
\]
with two factors and
\[
\bm\beta=(0.50,0.40,0.30,0.20,0.10)^\top
\]
with five factors. The treatment potential outcome was generated as
\[
Y_i(1)=Y_i(0)+\tau,
\]
where $\tau$ is the true additive treatment effect. Consequently, for the SIGA-R calculation at boundary $b$, the model-implied stratum-specific boundary deviations were
\[
d_s(b)=\tau-b,
\qquad s=1,\ldots,J.
\]
The adjusted score was obtained from a baseline-only linear working model containing the minimization-factor main effects but not the interaction term. The unadjusted and adjusted reference randomization tests used the same fixed, boundary-specific score vectors as the corresponding SIGA procedures.

Superiority was assessed using a two-sided level-$0.05$ test. Non-inferiority was assessed at the lower additive boundary $-0.20$ using a one-sided level-$0.025$ test. Equivalence was assessed over the interval $(-0.45,0.45)$ using two one-sided level-$0.05$ tests. Type I error was evaluated at the relevant constant-additive sharp boundary and, for equivalence, separately at the lower and upper limits. Fixed alternatives were selected in separate pilot simulations conducted before the final study so that the reference randomization test had approximately 85\% power, with values between 80\% and 90\% regarded as acceptable. The resulting prespecified alternatives are listed in Supplementary Table~1.

\subsection{Binary-outcome study}\label{sec:binary-sim}

The binary-outcome study used the same four minimization design--sample-size combinations and allocation settings as the continuous-outcome study. Unlike the homogeneous additive continuous model, the common-log-odds model below can produce different risk differences across joint strata because the baseline risks vary across strata. It therefore represents a practically motivated nonsharp setting without deliberately maximizing the difference between the two variance targets.

The conditional marginal potential-outcome probabilities were specified by
\[
\operatorname{logit}\Pbb\{Y_i(a)=1\mid\bm X_i\}
=
\alpha_0
+
\sum_{j=1}^F\gamma_jX_{ij}
+
\zeta X_{i1}X_{i2}
+
\theta a.
\]
For simulation, the observed outcome was drawn independently across participants as a Bernoulli variable with the probability corresponding to the realized assignment. This fully defines the observed-data law without requiring an otherwise unidentified coupling of the two binary potential outcomes. Supplementary Appendix B.1 shows that the variance combinations entering SIGA-S and SIGA-R depend only on the two conditional marginal outcome laws and not on that coupling.

The intercept $\alpha_0$ was calibrated so that the marginal control risk was $0.60$. For each scenario, $\theta$ was determined numerically by summing over the finite covariate distribution so that the resulting marginal risk difference agreed with its prespecified value to numerical tolerance. The first $F$ entries of
\[
(0.35,-0.25,0.20,-0.15,0.10)
\]
were used as the main-effect coefficients, and $\zeta=0.20$. The adjusted score was obtained by baseline-only linear residualization of $\bm W(b)$ on the minimization-factor main effects, with the interaction term omitted.

For the SIGA-R calculation, the model-implied boundary deviation in joint stratum $s$ was
\[
d_s(b)
=
\Pbb\{Y_i(1)=1\mid S_i=s\}
-
\Pbb\{Y_i(0)=1\mid S_i=s\}
-b.
\]
These values were calculated directly from the prespecified outcome-generating model rather than estimated separately in each outer trial.

Superiority was assessed using a two-sided level-$0.05$ test, and non-inferiority was assessed at the lower boundary $L=-0.10$ using a one-sided level-$0.025$ test. Equivalence was assessed using two one-sided level-$0.05$ tests. The symmetric equivalence limits were $(-0.21,0.21)$ for the two-factor design with 100 participants per group and $(-0.15,0.15)$ for the five-factor design with 200 participants per group; both larger designs used $(-0.10,0.10)$. For equivalence, type I error was evaluated separately at the lower and upper limits.

As in the continuous-outcome study, the design-specific treatment effects were selected in separate pilot calculations conducted before the final study to place the power of the reference randomization test in an informative range of approximately 80--90\%. For the two smaller binary-outcome designs, the equivalence limits were also selected in these pilot calculations for the same purpose. These simulation-specific limits were used only to compare the methods and should not be interpreted as clinically recommended equivalence margins. The prespecified treatment effects and equivalence limits are listed in Supplementary Table~2.

For unadjusted binary superiority at the boundary $b=0$, SIGA-S used the lattice-normal mixture refinement described in Supplementary Appendix D. Baseline-adjusted binary superiority and all binary non-inferiority and equivalence analyses used ordinary Gaussian SIGA-S tail probabilities. SIGA-R used ordinary Gaussian tail probabilities based on its randomization-calibrated variance; the lattice-normal refinement is not part of the pair-path correction.

\subsection{Operating characteristics and computation time}\label{sec:simulation-results}

Tables~\ref{tab:practical-unadjusted} and~\ref{tab:practical-adjusted} report all null-boundary and fixed-alternative rejection probabilities from the two practical simulation studies. In the sampling-targeted study, the largest absolute SIGA-S--reference-test difference across the 56 scenario configurations and both analyses was 0.42 percentage points; the maxima were 0.18 percentage points for null-boundary operating characteristics and 0.42 percentage points for power. In the independent randomization-targeted study, the largest absolute SIGA-R--reference-test difference was 0.32 percentage points; the corresponding maxima were 0.21 and 0.32 percentage points. Scenario-level mean variance ratios ranged from 1.000 to 1.042, and the numerical safeguard was not activated in any reported scenario--boundary--analysis combination. The sampling-targeted and randomization-targeted studies used independent outer simulations, as reflected by the two reference-test columns in each table.

\clearpage
\begingroup
\fontsize{8.5pt}{10.0pt}\selectfont
\setlength{\tabcolsep}{1.0pt}
\renewcommand{\arraystretch}{1.06}
\setlength{\LTleft}{\fill}
\setlength{\LTright}{\fill}
\begin{longtable}{@{}
>{\raggedright\arraybackslash}m{48pt}
>{\centering\arraybackslash}m{21pt}
>{\centering\arraybackslash}m{38pt}
>{\raggedright\arraybackslash}m{67pt}
>{\raggedright\arraybackslash}m{56pt}
>{\centering\arraybackslash}m{40pt}
*{4}{>{\centering\arraybackslash}m{35pt}}
>{\centering\arraybackslash}m{64pt}
@{}}
\caption{Complete operating characteristics for the unadjusted analysis in the practical simulation studies.}
\label{tab:practical-unadjusted}\\
\toprule
Outcome & $F$ & \shortstack{$n$/\\group} & Objective & Measure & Effect & \multicolumn{2}{c}{Sampling-targeted} & \multicolumn{3}{c}{Randomization-targeted} \\
\cmidrule(lr){7-8}\cmidrule(lr){9-11}
& & & & & & SIGA-S & RT & SIGA-R & RT & Mean $\widehat\rho_n$ \\
\midrule
\endfirsthead
\multicolumn{11}{c}{\tablename\ \thetable{} -- continued}\\
\toprule
Outcome & $F$ & \shortstack{$n$/\\group} & Objective & Measure & Effect & \multicolumn{2}{c}{Sampling-targeted} & \multicolumn{3}{c}{Randomization-targeted} \\
\cmidrule(lr){7-8}\cmidrule(lr){9-11}
& & & & & & SIGA-S & RT & SIGA-R & RT & Mean $\widehat\rho_n$ \\
\midrule
\endhead
\midrule
\multicolumn{11}{r}{Continued on next page}\\
\endfoot
\bottomrule
\endlastfoot
Continuous & 2 & 100 & Superiority & Type I error & +0.000 & 4.88 & 4.90 & 4.97 & 5.01 & 1.000 \\
Continuous & 2 & 100 & Superiority & Power & +0.430 & 85.20 & 85.17 & 84.92 & 85.17 & 1.010 \\
Continuous & 2 & 100 & Non-inferiority & Type I error & -0.200 & 2.44 & 2.46 & 2.50 & 2.54 & 1.000 \\
Continuous & 2 & 100 & Non-inferiority & Power & +0.230 & 85.14 & 85.06 & 84.85 & 85.06 & 1.010 \\
Continuous & 2 & 100 & Equivalence & Error at $L$ & -0.450 & 4.89 & 4.90 & 5.02 & 5.02 & 1.000/1.038 \\
Continuous & 2 & 100 & Equivalence & Error at $U$ & +0.450 & 4.99 & 4.95 & 4.96 & 4.92 & 1.038/1.000 \\
Continuous & 2 & 100 & Equivalence & Power & +0.000 & 86.96 & 86.91 & 86.37 & 86.48 & 1.011/1.011 \\
\cmidrule(lr){1-11}
Continuous & 2 & 500 & Superiority & Type I error & +0.000 & 4.89 & 4.88 & 5.08 & 5.09 & 1.000 \\
Continuous & 2 & 500 & Superiority & Power & +0.190 & 84.78 & 84.75 & 84.76 & 84.77 & 1.002 \\
Continuous & 2 & 500 & Non-inferiority & Type I error & -0.200 & 2.52 & 2.53 & 2.51 & 2.50 & 1.000 \\
Continuous & 2 & 500 & Non-inferiority & Power & -0.010 & 84.95 & 84.93 & 84.95 & 84.92 & 1.002 \\
Continuous & 2 & 500 & Equivalence & Error at $L$ & -0.450 & 5.02 & 5.01 & 4.98 & 4.98 & 1.000/1.035 \\
Continuous & 2 & 500 & Equivalence & Error at $U$ & +0.450 & 4.93 & 4.92 & 4.98 & 5.02 & 1.035/1.000 \\
Continuous & 2 & 500 & Equivalence & Power & +0.280 & 84.83 & 84.81 & 85.04 & 85.02 & 1.024/1.001 \\
\cmidrule(lr){1-11}
Binary & 2 & 100 & Superiority & Type I error & +0.000 & 4.16 & 4.29 & 4.08 & 4.25 & 1.000 \\
Binary & 2 & 100 & Superiority & Power & +0.200 & 86.11 & 86.21 & 86.00 & 86.32 & 1.011 \\
Binary & 2 & 100 & Non-inferiority & Type I error & -0.100 & 2.44 & 2.28 & 2.44 & 2.25 & 1.000 \\
Binary & 2 & 100 & Non-inferiority & Power & +0.100 & 84.08 & 83.85 & 83.90 & 83.85 & 1.010 \\
Binary & 2 & 100 & Equivalence & Error at $L$ & -0.210 & 4.77 & 4.68 & 4.74 & 4.65 & 1.000/1.035 \\
Binary & 2 & 100 & Equivalence & Error at $U$ & +0.210 & 4.90 & 4.89 & 4.97 & 5.00 & 1.042/1.000 \\
Binary & 2 & 100 & Equivalence & Power & +0.000 & 83.48 & 83.54 & 83.37 & 83.61 & 1.010/1.010 \\
\cmidrule(lr){1-11}
Binary & 2 & 500 & Superiority & Type I error & +0.000 & 4.55 & 4.54 & 4.68 & 4.67 & 1.000 \\
Binary & 2 & 500 & Superiority & Power & +0.100 & 91.07 & 91.06 & 91.09 & 91.19 & 1.002 \\
Binary & 2 & 500 & Non-inferiority & Type I error & -0.100 & 2.41 & 2.33 & 2.54 & 2.46 & 1.000 \\
Binary & 2 & 500 & Non-inferiority & Power & +0.000 & 89.68 & 89.69 & 89.79 & 89.87 & 1.002 \\
Binary & 2 & 500 & Equivalence & Error at $L$ & -0.100 & 5.04 & 4.89 & 4.96 & 4.79 & 1.000/1.008 \\
Binary & 2 & 500 & Equivalence & Error at $U$ & +0.100 & 5.05 & 4.95 & 5.04 & 4.97 & 1.009/1.000 \\
Binary & 2 & 500 & Equivalence & Power & +0.000 & 89.04 & 89.06 & 88.94 & 88.95 & 1.002/1.002 \\
\cmidrule(lr){1-11}
Continuous & 5 & 200 & Superiority & Type I error & +0.000 & 4.89 & 4.95 & 4.94 & 5.04 & 1.000 \\
Continuous & 5 & 200 & Superiority & Power & +0.300 & 84.18 & 84.32 & 84.13 & 84.37 & 1.004 \\
Continuous & 5 & 200 & Non-inferiority & Type I error & -0.200 & 2.48 & 2.52 & 2.45 & 2.51 & 1.000 \\
Continuous & 5 & 200 & Non-inferiority & Power & +0.100 & 84.28 & 84.36 & 84.26 & 84.46 & 1.004 \\
Continuous & 5 & 200 & Equivalence & Error at $L$ & -0.450 & 4.90 & 4.96 & 4.82 & 4.89 & 1.000/1.030 \\
Continuous & 5 & 200 & Equivalence & Error at $U$ & +0.450 & 4.89 & 4.95 & 4.94 & 5.02 & 1.030/1.000 \\
Continuous & 5 & 200 & Equivalence & Power & +0.180 & 84.67 & 84.75 & 84.74 & 84.90 & 1.016/1.003 \\
\cmidrule(lr){1-11}
Continuous & 5 & 1000 & Superiority & Type I error & +0.000 & 5.02 & 5.01 & 5.12 & 5.11 & 1.000 \\
Continuous & 5 & 1000 & Superiority & Power & +0.135 & 85.11 & 85.13 & 85.22 & 85.23 & 1.001 \\
Continuous & 5 & 1000 & Non-inferiority & Type I error & -0.200 & 2.50 & 2.50 & 2.56 & 2.57 & 1.000 \\
Continuous & 5 & 1000 & Non-inferiority & Power & -0.065 & 85.43 & 85.38 & 85.15 & 85.16 & 1.001 \\
Continuous & 5 & 1000 & Equivalence & Error at $L$ & -0.450 & 4.95 & 4.99 & 4.95 & 4.95 & 1.000/1.026 \\
Continuous & 5 & 1000 & Equivalence & Error at $U$ & +0.450 & 5.09 & 5.12 & 4.87 & 4.92 & 1.026/1.000 \\
Continuous & 5 & 1000 & Equivalence & Power & +0.330 & 84.75 & 84.78 & 84.75 & 84.80 & 1.020/1.001 \\
\cmidrule(lr){1-11}
Binary & 5 & 200 & Superiority & Type I error & +0.000 & 4.58 & 4.58 & 4.47 & 4.45 & 1.000 \\
Binary & 5 & 200 & Superiority & Power & +0.140 & 83.99 & 84.03 & 83.93 & 84.06 & 1.004 \\
Binary & 5 & 200 & Non-inferiority & Type I error & -0.100 & 2.38 & 2.38 & 2.31 & 2.30 & 1.000 \\
Binary & 5 & 200 & Non-inferiority & Power & +0.050 & 87.43 & 87.33 & 87.20 & 87.19 & 1.004 \\
Binary & 5 & 200 & Equivalence & Error at $L$ & -0.150 & 4.69 & 4.79 & 4.65 & 4.76 & 1.000/1.015 \\
Binary & 5 & 200 & Equivalence & Error at $U$ & +0.150 & 4.88 & 4.88 & 5.03 & 4.99 & 1.017/1.000 \\
Binary & 5 & 200 & Equivalence & Power & +0.000 & 84.89 & 84.85 & 84.64 & 84.74 & 1.004/1.004 \\
\cmidrule(lr){1-11}
Binary & 5 & 1000 & Superiority & Type I error & +0.000 & 4.67 & 4.70 & 4.64 & 4.66 & 1.000 \\
Binary & 5 & 1000 & Superiority & Power & +0.070 & 90.05 & 90.03 & 90.23 & 90.19 & 1.001 \\
Binary & 5 & 1000 & Non-inferiority & Type I error & -0.100 & 2.45 & 2.47 & 2.40 & 2.42 & 1.000 \\
Binary & 5 & 1000 & Non-inferiority & Power & -0.030 & 89.27 & 89.15 & 89.35 & 89.24 & 1.001 \\
Binary & 5 & 1000 & Equivalence & Error at $L$ & -0.100 & 4.64 & 4.71 & 4.79 & 4.84 & 1.000/1.006 \\
Binary & 5 & 1000 & Equivalence & Error at $U$ & +0.100 & 4.98 & 4.95 & 5.00 & 4.97 & 1.007/1.000 \\
Binary & 5 & 1000 & Equivalence & Power & +0.040 & 87.22 & 87.22 & 87.38 & 87.35 & 1.003/1.001 \\
\end{longtable}
\noindent{\footnotesize Rejection probabilities are percentages. The sampling-targeted and randomization-targeted studies used independent outer simulations, so the two RT columns need not be numerically identical. For equivalence, $L$ and $U$ denote the lower and upper limits, and two variance ratios are reported in lower-boundary/upper-boundary order.}
\endgroup

\begingroup
\fontsize{8.5pt}{10.0pt}\selectfont
\setlength{\tabcolsep}{1.0pt}
\renewcommand{\arraystretch}{1.06}
\setlength{\LTleft}{\fill}
\setlength{\LTright}{\fill}
\begin{longtable}{@{}
>{\raggedright\arraybackslash}m{48pt}
>{\centering\arraybackslash}m{21pt}
>{\centering\arraybackslash}m{38pt}
>{\raggedright\arraybackslash}m{67pt}
>{\raggedright\arraybackslash}m{56pt}
>{\centering\arraybackslash}m{40pt}
*{4}{>{\centering\arraybackslash}m{35pt}}
>{\centering\arraybackslash}m{64pt}
@{}}
\caption{Complete operating characteristics for the baseline-adjusted analysis in the practical simulation studies.}
\label{tab:practical-adjusted}\\
\toprule
Outcome & $F$ & \shortstack{$n$/\\group} & Objective & Measure & Effect & \multicolumn{2}{c}{Sampling-targeted} & \multicolumn{3}{c}{Randomization-targeted} \\
\cmidrule(lr){7-8}\cmidrule(lr){9-11}
& & & & & & SIGA-S & RT & SIGA-R & RT & Mean $\widehat\rho_n$ \\
\midrule
\endfirsthead
\multicolumn{11}{c}{\tablename\ \thetable{} -- continued}\\
\toprule
Outcome & $F$ & \shortstack{$n$/\\group} & Objective & Measure & Effect & \multicolumn{2}{c}{Sampling-targeted} & \multicolumn{3}{c}{Randomization-targeted} \\
\cmidrule(lr){7-8}\cmidrule(lr){9-11}
& & & & & & SIGA-S & RT & SIGA-R & RT & Mean $\widehat\rho_n$ \\
\midrule
\endhead
\midrule
\multicolumn{11}{r}{Continued on next page}\\
\endfoot
\bottomrule
\endlastfoot
Continuous & 2 & 100 & Superiority & Type I error & +0.000 & 4.87 & 4.88 & 4.93 & 4.98 & 1.000 \\
Continuous & 2 & 100 & Superiority & Power & +0.430 & 85.50 & 85.44 & 85.20 & 85.40 & 1.010 \\
Continuous & 2 & 100 & Non-inferiority & Type I error & -0.200 & 2.44 & 2.44 & 2.51 & 2.51 & 1.000 \\
Continuous & 2 & 100 & Non-inferiority & Power & +0.230 & 85.41 & 85.37 & 85.08 & 85.29 & 1.010 \\
Continuous & 2 & 100 & Equivalence & Error at $L$ & -0.450 & 4.92 & 4.94 & 5.01 & 5.00 & 1.000/1.038 \\
Continuous & 2 & 100 & Equivalence & Error at $U$ & +0.450 & 4.96 & 4.94 & 4.98 & 4.94 & 1.038/1.000 \\
Continuous & 2 & 100 & Equivalence & Power & +0.000 & 87.28 & 87.17 & 86.68 & 86.83 & 1.011/1.011 \\
\cmidrule(lr){1-11}
Continuous & 2 & 500 & Superiority & Type I error & +0.000 & 4.87 & 4.88 & 5.08 & 5.07 & 1.000 \\
Continuous & 2 & 500 & Superiority & Power & +0.190 & 84.80 & 84.78 & 84.87 & 84.85 & 1.002 \\
Continuous & 2 & 500 & Non-inferiority & Type I error & -0.200 & 2.52 & 2.54 & 2.50 & 2.51 & 1.000 \\
Continuous & 2 & 500 & Non-inferiority & Power & -0.010 & 84.99 & 84.95 & 84.97 & 84.96 & 1.002 \\
Continuous & 2 & 500 & Equivalence & Error at $L$ & -0.450 & 5.04 & 5.00 & 4.96 & 4.97 & 1.000/1.035 \\
Continuous & 2 & 500 & Equivalence & Error at $U$ & +0.450 & 4.90 & 4.92 & 4.98 & 5.00 & 1.035/1.000 \\
Continuous & 2 & 500 & Equivalence & Power & +0.280 & 84.87 & 84.85 & 85.09 & 85.04 & 1.025/1.001 \\
\cmidrule(lr){1-11}
Binary & 2 & 100 & Superiority & Type I error & +0.000 & 5.10 & 4.92 & 5.02 & 4.81 & 1.000 \\
Binary & 2 & 100 & Superiority & Power & +0.200 & 87.92 & 87.50 & 87.83 & 87.62 & 1.011 \\
Binary & 2 & 100 & Non-inferiority & Type I error & -0.100 & 2.37 & 2.32 & 2.36 & 2.28 & 1.000 \\
Binary & 2 & 100 & Non-inferiority & Power & +0.100 & 84.13 & 84.06 & 83.90 & 84.00 & 1.010 \\
Binary & 2 & 100 & Equivalence & Error at $L$ & -0.210 & 4.76 & 4.66 & 4.73 & 4.62 & 1.000/1.035 \\
Binary & 2 & 100 & Equivalence & Error at $U$ & +0.210 & 4.88 & 4.86 & 4.96 & 4.99 & 1.042/1.000 \\
Binary & 2 & 100 & Equivalence & Power & +0.000 & 83.64 & 83.58 & 83.48 & 83.65 & 1.010/1.010 \\
\cmidrule(lr){1-11}
Binary & 2 & 500 & Superiority & Type I error & +0.000 & 4.85 & 4.83 & 5.00 & 4.97 & 1.000 \\
Binary & 2 & 500 & Superiority & Power & +0.100 & 91.62 & 91.51 & 91.66 & 91.61 & 1.002 \\
Binary & 2 & 500 & Non-inferiority & Type I error & -0.100 & 2.39 & 2.35 & 2.51 & 2.47 & 1.000 \\
Binary & 2 & 500 & Non-inferiority & Power & +0.000 & 89.72 & 89.73 & 89.83 & 89.93 & 1.002 \\
Binary & 2 & 500 & Equivalence & Error at $L$ & -0.100 & 5.00 & 4.92 & 4.91 & 4.82 & 1.000/1.008 \\
Binary & 2 & 500 & Equivalence & Error at $U$ & +0.100 & 5.06 & 5.00 & 5.03 & 5.00 & 1.009/1.000 \\
Binary & 2 & 500 & Equivalence & Power & +0.000 & 89.07 & 89.11 & 88.98 & 89.00 & 1.002/1.002 \\
\cmidrule(lr){1-11}
Continuous & 5 & 200 & Superiority & Type I error & +0.000 & 4.99 & 4.97 & 5.00 & 4.99 & 1.000 \\
Continuous & 5 & 200 & Superiority & Power & +0.300 & 84.71 & 84.65 & 84.73 & 84.79 & 1.004 \\
Continuous & 5 & 200 & Non-inferiority & Type I error & -0.200 & 2.53 & 2.52 & 2.51 & 2.54 & 1.000 \\
Continuous & 5 & 200 & Non-inferiority & Power & +0.100 & 84.79 & 84.75 & 84.79 & 84.81 & 1.004 \\
Continuous & 5 & 200 & Equivalence & Error at $L$ & -0.450 & 4.98 & 4.99 & 4.92 & 4.91 & 1.000/1.030 \\
Continuous & 5 & 200 & Equivalence & Error at $U$ & +0.450 & 4.95 & 4.97 & 5.07 & 5.08 & 1.030/1.000 \\
Continuous & 5 & 200 & Equivalence & Power & +0.180 & 85.14 & 85.06 & 85.11 & 85.09 & 1.016/1.003 \\
\cmidrule(lr){1-11}
Continuous & 5 & 1000 & Superiority & Type I error & +0.000 & 5.01 & 5.00 & 5.11 & 5.11 & 1.000 \\
Continuous & 5 & 1000 & Superiority & Power & +0.135 & 85.20 & 85.17 & 85.37 & 85.33 & 1.001 \\
Continuous & 5 & 1000 & Non-inferiority & Type I error & -0.200 & 2.54 & 2.52 & 2.57 & 2.59 & 1.000 \\
Continuous & 5 & 1000 & Non-inferiority & Power & -0.065 & 85.52 & 85.44 & 85.28 & 85.27 & 1.001 \\
Continuous & 5 & 1000 & Equivalence & Error at $L$ & -0.450 & 4.98 & 4.98 & 4.97 & 4.94 & 1.000/1.026 \\
Continuous & 5 & 1000 & Equivalence & Error at $U$ & +0.450 & 5.12 & 5.11 & 4.92 & 4.93 & 1.026/1.000 \\
Continuous & 5 & 1000 & Equivalence & Power & +0.330 & 84.84 & 84.82 & 84.87 & 84.83 & 1.020/1.001 \\
\cmidrule(lr){1-11}
Binary & 5 & 200 & Superiority & Type I error & +0.000 & 5.08 & 5.10 & 4.97 & 5.00 & 1.000 \\
Binary & 5 & 200 & Superiority & Power & +0.140 & 85.32 & 85.24 & 85.27 & 85.24 & 1.004 \\
Binary & 5 & 200 & Non-inferiority & Type I error & -0.100 & 2.41 & 2.42 & 2.33 & 2.33 & 1.000 \\
Binary & 5 & 200 & Non-inferiority & Power & +0.050 & 87.50 & 87.43 & 87.30 & 87.32 & 1.004 \\
Binary & 5 & 200 & Equivalence & Error at $L$ & -0.150 & 4.84 & 4.85 & 4.82 & 4.83 & 1.000/1.015 \\
Binary & 5 & 200 & Equivalence & Error at $U$ & +0.150 & 4.90 & 4.91 & 5.06 & 5.03 & 1.017/1.000 \\
Binary & 5 & 200 & Equivalence & Power & +0.000 & 85.06 & 84.97 & 84.90 & 84.88 & 1.004/1.004 \\
\cmidrule(lr){1-11}
Binary & 5 & 1000 & Superiority & Type I error & +0.000 & 4.92 & 4.92 & 4.91 & 4.88 & 1.000 \\
Binary & 5 & 1000 & Superiority & Power & +0.070 & 90.45 & 90.42 & 90.61 & 90.56 & 1.001 \\
Binary & 5 & 1000 & Non-inferiority & Type I error & -0.100 & 2.48 & 2.50 & 2.43 & 2.44 & 1.000 \\
Binary & 5 & 1000 & Non-inferiority & Power & -0.030 & 89.22 & 89.19 & 89.30 & 89.28 & 1.001 \\
Binary & 5 & 1000 & Equivalence & Error at $L$ & -0.100 & 4.72 & 4.75 & 4.86 & 4.86 & 1.000/1.006 \\
Binary & 5 & 1000 & Equivalence & Error at $U$ & +0.100 & 4.97 & 4.97 & 5.00 & 4.99 & 1.007/1.000 \\
Binary & 5 & 1000 & Equivalence & Power & +0.040 & 87.25 & 87.26 & 87.41 & 87.43 & 1.003/1.001 \\
\end{longtable}
\noindent{\footnotesize Rejection probabilities are percentages. The sampling-targeted and randomization-targeted studies used independent outer simulations, so the two RT columns need not be numerically identical. For equivalence, $L$ and $U$ denote the lower and upper limits, and two variance ratios are reported in lower-boundary/upper-boundary order.}
\endgroup

The timing comparison directly addressed the computational motivation for SIGA. It was conducted on a single Apple M3 Ultra workstation using one CPU core per process. Because it preceded the final paired implementation, it compared SIGA-S with the reference randomization test; the trial-inspired illustration below provides a same-machine comparison of all three procedures. Table~\ref{tab:main-computation-times} reports the complete projected cumulative single-core runtimes for 100,000 analyses. Across the practical simulation settings, SIGA-S required 0.050--2.259 minutes, compared with 113.04--2625.76 minutes for the reference randomization test. Table~\ref{tab:main-calibration-times} reports the one-time allocation-only calibration times, which ranged from 2.821 to 47.545 seconds. Thus, reusing the allocation-only calibration reduced computation by several orders of magnitude relative to nested rerandomization. The timing protocol is detailed in Supplementary Appendix F.

\begingroup
\fontsize{9pt}{10.5pt}\selectfont
\setlength{\tabcolsep}{1.3pt}
\renewcommand{\arraystretch}{1.06}
\setlength{\LTleft}{\fill}
\setlength{\LTright}{\fill}
\begin{longtable}{@{}
>{\raggedright\arraybackslash}m{54pt}
>{\centering\arraybackslash}m{22pt}
>{\centering\arraybackslash}m{42pt}
>{\raggedright\arraybackslash}m{62pt}
>{\raggedright\arraybackslash}m{84pt}
>{\centering\arraybackslash}m{103pt}
>{\centering\arraybackslash}m{103pt}
@{}}
\caption{Projected single-core computation times for 100,000 analyses in the practical simulation settings.}
\label{tab:main-computation-times}\\
\toprule
Outcome & $F$ & \shortstack{$n$/\\group} & Analysis & Objective & \shortstack{SIGA-S projected\\time (min)} & \shortstack{RT projected\\time (min)} \\
\midrule
\endfirsthead
\multicolumn{7}{c}{\tablename\ \thetable{} -- continued}\\
\toprule
Outcome & $F$ & \shortstack{$n$/\\group} & Analysis & Objective & \shortstack{SIGA-S projected\\time (min)} & \shortstack{RT projected\\time (min)} \\
\midrule
\endhead
\midrule
\multicolumn{7}{r}{Continued on next page}\\
\endfoot
\bottomrule
\endlastfoot
Continuous & 2 & 100 & Unadjusted & Superiority & 0.050 & 113.54 \\
Continuous & 2 & 100 & Adjusted & Superiority & 0.051 & 113.04 \\
Continuous & 2 & 100 & Unadjusted & Non-inferiority & 0.051 & 114.04 \\
Continuous & 2 & 100 & Adjusted & Non-inferiority & 0.050 & 114.22 \\
Continuous & 2 & 100 & Unadjusted & Equivalence & 0.098 & 120.76 \\
Continuous & 2 & 100 & Adjusted & Equivalence & 0.099 & 121.50 \\
\cmidrule(lr){1-7}
Continuous & 2 & 500 & Unadjusted & Superiority & 0.103 & 559.54 \\
Continuous & 2 & 500 & Adjusted & Superiority & 0.105 & 561.20 \\
Continuous & 2 & 500 & Unadjusted & Non-inferiority & 0.105 & 557.00 \\
Continuous & 2 & 500 & Adjusted & Non-inferiority & 0.102 & 559.78 \\
Continuous & 2 & 500 & Unadjusted & Equivalence & 0.205 & 590.24 \\
Continuous & 2 & 500 & Adjusted & Equivalence & 0.206 & 591.31 \\
\cmidrule(lr){1-7}
Binary & 2 & 100 & Unadjusted & Superiority & 0.122 & 114.24 \\
Binary & 2 & 100 & Adjusted & Superiority & 0.050 & 114.26 \\
Binary & 2 & 100 & Unadjusted & Non-inferiority & 0.051 & 114.39 \\
Binary & 2 & 100 & Adjusted & Non-inferiority & 0.050 & 114.26 \\
Binary & 2 & 100 & Unadjusted & Equivalence & 0.098 & 121.41 \\
Binary & 2 & 100 & Adjusted & Equivalence & 0.099 & 121.39 \\
\cmidrule(lr){1-7}
Binary & 2 & 500 & Unadjusted & Superiority & 2.259 & 775.46 \\
Binary & 2 & 500 & Adjusted & Superiority & 1.574 & 762.43 \\
Binary & 2 & 500 & Unadjusted & Non-inferiority & 0.296 & 779.28 \\
Binary & 2 & 500 & Adjusted & Non-inferiority & 0.204 & 758.11 \\
Binary & 2 & 500 & Unadjusted & Equivalence & 0.444 & 894.15 \\
Binary & 2 & 500 & Adjusted & Equivalence & 0.352 & 802.63 \\
\cmidrule(lr){1-7}
Continuous & 5 & 200 & Unadjusted & Superiority & 0.096 & 327.26 \\
Continuous & 5 & 200 & Adjusted & Superiority & 0.097 & 327.54 \\
Continuous & 5 & 200 & Unadjusted & Non-inferiority & 0.096 & 328.00 \\
Continuous & 5 & 200 & Adjusted & Non-inferiority & 0.096 & 327.61 \\
Continuous & 5 & 200 & Unadjusted & Equivalence & 0.192 & 342.31 \\
Continuous & 5 & 200 & Adjusted & Equivalence & 0.191 & 342.83 \\
\cmidrule(lr){1-7}
Continuous & 5 & 1000 & Unadjusted & Superiority & 0.290 & 1618.13 \\
Continuous & 5 & 1000 & Adjusted & Superiority & 0.290 & 1619.24 \\
Continuous & 5 & 1000 & Unadjusted & Non-inferiority & 0.288 & 1615.09 \\
Continuous & 5 & 1000 & Adjusted & Non-inferiority & 0.285 & 1616.06 \\
Continuous & 5 & 1000 & Unadjusted & Equivalence & 0.573 & 1683.11 \\
Continuous & 5 & 1000 & Adjusted & Equivalence & 0.581 & 1685.70 \\
\cmidrule(lr){1-7}
Binary & 5 & 200 & Unadjusted & Superiority & 0.220 & 329.20 \\
Binary & 5 & 200 & Adjusted & Superiority & 0.095 & 329.22 \\
Binary & 5 & 200 & Unadjusted & Non-inferiority & 0.095 & 330.28 \\
Binary & 5 & 200 & Adjusted & Non-inferiority & 0.094 & 329.89 \\
Binary & 5 & 200 & Unadjusted & Equivalence & 0.189 & 343.78 \\
Binary & 5 & 200 & Adjusted & Equivalence & 0.188 & 344.41 \\
\cmidrule(lr){1-7}
Binary & 5 & 1000 & Unadjusted & Superiority & 1.074 & 2543.70 \\
Binary & 5 & 1000 & Adjusted & Superiority & 0.519 & 2309.76 \\
Binary & 5 & 1000 & Unadjusted & Non-inferiority & 0.611 & 2474.69 \\
Binary & 5 & 1000 & Adjusted & Non-inferiority & 0.463 & 2348.11 \\
Binary & 5 & 1000 & Unadjusted & Equivalence & 0.778 & 2625.76 \\
Binary & 5 & 1000 & Adjusted & Equivalence & 0.796 & 2609.09 \\
\end{longtable}
\endgroup

\begin{table}[!htbp]
\centering
\caption{Single-core computation times for one reusable allocation-only SIGA-S calibration.}
\label{tab:main-calibration-times}
\small
\setlength{\tabcolsep}{3pt}
\renewcommand{\arraystretch}{1.08}
\begin{threeparttable}
\begin{tabular}{rrrrrrrr}
\toprule
Factors & $n$/group & Total $n$ & $B_0$ & Run 1 (s) & Run 2 (s) & Run 3 (s) & Mean (s) \\
\midrule
2 & 100  & 200  & 100,000 & 2.851  & 2.821  & 2.792  & 2.821 \\
2 & 500  & 1000 & 100,000 & 13.844 & 13.829 & 13.809 & 13.827 \\
5 & 200  & 400  & 100,000 & 9.604  & 9.575  & 9.607  & 9.595 \\
5 & 1000 & 2000 & 100,000 & 47.527 & 47.530 & 47.579 & 47.545 \\
\bottomrule
\end{tabular}
\begin{tablenotes}[flushleft]
\footnotesize
\item Runs 1--3 are independent repetitions of the timing benchmark for the same one-time calibration procedure, with $B_0=100{,}000$ allocation-only paths in each run. Only one allocation-only calibration is required in practice for each combination of the minimization design, factor distribution, and total sample size. The resulting calibration can be reused across outcome types, testing objectives, null boundaries, treatment-effect assumptions, and prespecified score constructions.
\end{tablenotes}
\end{threeparttable}
\end{table}

\FloatBarrier

\paragraph{Targeted pair-path sensitivity analysis.}
The complete sensitivity-analysis results are reported in Supplementary Appendix G and Supplementary Table~4. In the practically heterogeneous scenarios S01--S08, the mean diagnostic ratio $\widehat\rho_n$ ranged from 1.000 to 1.004 and the largest absolute difference between either SIGA procedure and the reference randomization test was 0.14 percentage points. In the strong continuous scenarios S09 and S11, the mean ratio was 1.193--1.194; SIGA-R was within 0.04 percentage points of the reference test, whereas the largest SIGA-S difference was 0.90 percentage points. In the strong binary scenarios S10 and S12, the mean ratio was 1.113--1.115 and the largest SIGA-R difference was 0.24 percentage points. The numerical safeguard was not activated in any of the 24 scenario--analysis combinations. These results support the predicted variance separation and show that the pair-path correction becomes consequential when the conditional randomization variance is materially larger than the repeated-sampling variance.

\section{Trial-inspired power illustration}\label{sec:case-study}

We illustrate the design-stage use of the SIGA framework using the published planning characteristics of SWIFT DIRECT, a multicentre randomized non-inferiority trial comparing direct mechanical thrombectomy with intravenous alteplase followed by thrombectomy in patients with acute ischaemic stroke \citep{FischerEtAl2022Protocol,FischerEtAl2022Trial}. The binary primary endpoint was functional independence, defined as a modified Rankin Scale score of 0--2 at 90 days. The protocol planned a total of 404 participants and reported 80\% power under an assumed response probability of 0.622 in each group, an absolute non-inferiority margin of 0.12 on the treatment-minus-control risk-difference scale, and a one-sided significance level of 0.05. The treatment coefficient in the outcome model is set to zero, so that both marginal response probabilities are 0.622 and the true risk difference is zero.

The illustration retains the five dichotomized factors reported for the trial's minimization and covariate-adjusted analyses: baseline National Institutes of Health Stroke Scale score ($\le17$ versus $>17$), age ($<70$ versus $\ge70$ years), occlusion location (M1 only versus internal-carotid-artery-related occlusion), tandem-lesion status, and Alberta Stroke Program Early CT Score (4--7 versus 8--10). The working prevalences of the adverse categories are 0.466, 0.592, 0.287, 0.154 and 0.300, respectively. The NIHSS and age prevalences are approximated from the pooled medians and interquartile ranges using truncated-normal working distributions. The occlusion and tandem-lesion prevalences are based on pooled counts, whereas the ASPECTS prevalence is based on a discrete working distribution chosen to reproduce the reported pooled median of 8 and interquartile range of 7--9. Because the joint distribution of the five factors was not reported, they are generated independently. No correlation structure is imposed in this trial-inspired reconstruction without participant-level information. The illustration is therefore neither a participant-level reanalysis nor an exact reconstruction of the completed trial.

To make the minimization factors prognostic, binary outcomes are generated from a logistic model containing their five main effects. The coefficients for the adverse categories are set to $(-0.55,-0.25,-0.40,-0.25,-0.60)$ for NIHSS, age, occlusion location, tandem lesion and ASPECTS, respectively. These values represent moderate working prognostic associations and are treated as design assumptions because factor-specific outcome effects were not available from the aggregate reports. No interaction terms are included. The intercept is calibrated so that the marginal response probability is 0.622, and the treatment coefficient is set exactly to zero. The treatment and control groups therefore have identical conditional and marginal response probabilities.

The published trial used deterministic minimization, but the aggregate publications do not provide the participant-level information required to reconstruct the realized allocation process. To illustrate prospective power evaluation under a fully specified stochastic randomization design, treatment is assigned using the 1:1 label-symmetric biased-coin minimization rule in \eqref{eq:min-score} with $p_{\rm bc}=0.80$. This rule is a prespecified randomized analogue used for the illustration and is not intended to reproduce the completed trial's allocation process exactly. Non-inferiority at the lower boundary $-0.12$ is evaluated using both the unadjusted score and baseline-only linear residualization on the five factor main effects. SIGA-S, SIGA-R and the reference randomization test are evaluated on common outer trials. At this nonsharp binary risk-difference boundary, the fixed-score reference test has no compatible nonzero constant-additive binary sharp-null interpretation. SIGA-S targets the boundary-valid marginal risk-difference test, whereas SIGA-R targets the conditional distribution of the fixed-score reference test; their agreement is therefore assessed empirically rather than assumed.

The final calculation used 100,000 common outer trials and 4,999 regenerated paths for each reference test. SIGA-S used a one-path calibration based on 100,000 allocation-only replicates. For SIGA-R, an additional 100,000-replicate three-path calibration supplied the pair-path covariance while retaining the one-path covariance calibration used by SIGA-S. At total sample size 404, the unadjusted rejection probabilities were 81.48\% for SIGA-S, 81.44\% for SIGA-R and 81.28\% for the reference test. The corresponding baseline-adjusted values were 81.53\%, 81.48\% and 81.50\%, respectively. The mean diagnostic ratio was 1.003 for both score analyses, indicating close alignment of the repeated-sampling and conditional randomization variance targets in this setting. Thus, under the stated trial-inspired model, all three procedures gave similar power. The largest pairwise difference was 0.20 percentage points, and the SIGA-R--reference-test differences were 0.16 percentage points for the unadjusted analysis and 0.02 percentage points for the baseline-adjusted analysis.

On the same Apple M3 Ultra benchmark machine, the one-time calibration and analysis of all 100,000 trials required 0.98 minutes for SIGA-S and 0.85 minutes for SIGA-R, compared with 406.36 minutes for nested rerandomization. The SIGA-R calibration includes both one-path and pair-path covariance quantities. All method-specific times exclude common data generation and score construction.

\begin{table}[!htbp]
\centering
\caption{Power and computation time in the SWIFT DIRECT-inspired simulation.}
\label{tab:swift-direct-siga-summary}
\begin{threeparttable}
\small
\setlength{\tabcolsep}{5.0pt}
\renewcommand{\arraystretch}{1.08}
\begin{tabular}{lrrr}
\toprule
 & SIGA-S & SIGA-R & RT \\
\midrule
\multicolumn{4}{l}{\textit{Estimated non-inferiority power (\%)}}\\
Unadjusted score & 81.48 & 81.44 & 81.28 \\
Baseline-adjusted score & 81.53 & 81.48 & 81.50 \\
\addlinespace
\multicolumn{4}{l}{\textit{Computation time for all 100,000 trials (min)}}\\
One-time allocation calibration & 0.21 & 0.49 & -- \\
Complete analyses & 0.77 & 0.36 & 406.36 \\
Total method-specific time & 0.98 & 0.85 & 406.36 \\
\bottomrule
\end{tabular}
\begin{tablenotes}[flushleft]
\footnotesize
\item RT denotes the fixed-score reference randomization test. All procedures used common outer trials. SIGA-S targets the boundary-valid marginal-risk-difference test, whereas SIGA-R targets the fixed-score reference randomization test. Data generation and score construction are excluded from the timing rows.
\end{tablenotes}
\end{threeparttable}
\end{table}

\section{Discussion}\label{sec:discussion}

This work transforms design-stage power and sample-size evaluation under biased-coin minimization from a computationally intensive nested rerandomization task into a reusable calibration problem. The SIGA framework makes the inferential target explicit by separating two distributions that can differ at a nonsharp boundary. SIGA-S calibrates the repeated-sampling variance of a marginal mean- or risk-difference statistic and supports boundary-valid inference for the marginal average effect. SIGA-R adds the pair-path covariance required to reproduce the conditional distribution of a prespecified fixed-score reference randomization test. Because the allocation-only calibrations can be reused across outcome-generating trials and design scenarios, both procedures remove the dominant computational burden without obscuring the quantity being approximated. The principal practical advance is therefore not only a substantial reduction in computation, but also a transparent framework for matching the design-stage calculation to the intended inferential target.

Methodologically, the framework is built on three connected developments. First, the fixed-score statistic is decomposed exactly into a joint-stratum imbalance component and an orthogonal within-stratum component, thereby isolating the part of the statistic affected by the minimization design. Second, the repeated-sampling and conditional randomization variances are derived separately, and their difference is expressed as a quadratic form involving the stratum-specific boundary deviations and the pair-path covariance. This variance-gap identity characterizes when the two targets agree and explains how the conditional variance of the fixed-score reference test changes when stratum-specific average effects differ from the tested boundary. Third, the required covariance quantities are estimated using reusable one-path and paired-path allocation-only calibrations, while the variance-ratio diagnostic quantifies the practical importance of the pair-path correction. The resulting framework applies to continuous and binary outcomes on their numerical scales, accommodates fixed-dimensional prespecified baseline-only adjustment, and does not require an unidentified within-participant coupling of binary potential outcomes. Its asymptotic results are established under the allocation-limit, moment and nonsingularity conditions stated in the Methods and Supplementary Material.

The numerical studies were designed to evaluate each procedure in relation to the distribution it is intended to represent. In two independent simulation studies using the same 56 practically motivated scenario configurations, the largest absolute differences from the corresponding reference-test rejection probabilities were 0.42 percentage points for SIGA-S and 0.32 percentage points for SIGA-R. These findings do not rank the procedures, because they address different inferential targets and were evaluated using independent outer simulations. In the practical scenarios, the mean randomization-to-sampling variance ratios ranged from 1.000 to 1.042, explaining why the two calibrations generally produced similar operating characteristics. The targeted pair-path sensitivity analysis in Supplementary Appendix G deliberately increased the discrepancy through stronger biased-coin allocation and maximum-ratio treatment-effect heterogeneity. When the mean ratio reached approximately 1.19 in the continuous settings, the reference randomization test became conservative relative to the repeated-sampling target: SIGA-R tracked this shift to within 0.04 percentage points, whereas SIGA-S remained near the nominal level and differed by as much as 0.90 percentage points. The binary sensitivity settings showed the same direction of change, although the remaining SIGA-R--reference-test difference reached 0.24 percentage points, indicating that the first-order variance correction does not remove all finite-sample or discrete-tail discrepancies. Thus, the sensitivity analysis reinforces the intended interpretation of the two procedures: SIGA-S remains the boundary-valid method for marginal average-effect inference, whereas SIGA-R is the appropriate design-stage surrogate when the prespecified target is the fixed-score reference randomization test and the pair-path variance contribution is non-negligible.

The SWIFT DIRECT-inspired illustration further demonstrates the practical value of the framework for prospective design calculations. Under the stated working model, the power estimates from SIGA-S, SIGA-R and the reference randomization test differed by at most 0.20 percentage points. The complete calculations for 100,000 simulated trials required 0.98 minutes for SIGA-S and 0.85 minutes for SIGA-R, compared with 406.36 minutes for nested rerandomization. Thus, the reusable calibrations preserved close agreement with the intended reference analysis while making randomization-aware power evaluation operationally feasible.

For practical implementation, the choice between SIGA-S and SIGA-R should be determined by the intended final analysis rather than by which procedure gives the more favourable power estimate. We recommend SIGA-S when the design objective is valid testing and power or sample-size evaluation for a marginal average effect, including marginal mean and risk differences. We recommend SIGA-R when the final analysis has been prespecified as a fixed-score reference randomization test and the design-stage calculation is intended to reproduce that test's rejection probability. The variance-ratio diagnostic should accompany the SIGA-R calculation: values close to one indicate that the pair-path correction has little practical effect, whereas departures from one quantify the additional conditional-variance contribution captured by SIGA-R. Thus, the proposed framework provides a simple analysis-driven workflow: first specify the inferential target, then apply the corresponding calibration, and use the diagnostic ratio to interpret the difference between the repeated-sampling and conditional randomization targets. The exact score projection, explicit separation of inferential targets and reusable allocation-only calibrations provide a coherent, accurate and computationally efficient basis for power and sample-size evaluation under biased-coin minimization.

\section*{Acknowledgments}
This work was supported by the Japan Society for the Promotion of Science (JSPS) KAKENHI (Grant Number JP26K21185).

The author declares no conflicts of interest.

\section*{Data availability}
No individual participant data were analyzed. The SWIFT DIRECT illustration is a simulation based on published aggregate planning characteristics. The R code is available on GitHub at \CodeRepository \citep{KojimaSIGA2026}, and the exact code snapshot accompanying this submission is provided as Supplementary Software.

\section*{Supplementary Material}
Supplementary Appendices A--G, Supplementary Tables 1--4 and the accompanying software archive are provided as Supplementary Material.

\begingroup\small
\bibliographystyle{plainnat}
\bibliography{main}
\endgroup
\end{document}